\documentclass[12pt]{revtex4} \usepackage[dvips]{graphicx,color}
\def\bea{\begin{eqnarray}}
\def\eea{\end{eqnarray}}
\def\be{\begin{equation}}
\def\ee{\end{equation}}
\def\nn{\nonumber}

\def\i{\imath}

\def\d{\delta}

\def\g{\gamma}
\def\G{\Gamma}

\def\b{\beta}\def\th{\theta}

\begin{document}
{\flushright{\tiny{UNB-Technical-Report 05/10}}}
\title{Generalised coherent states and combinatorics of horizon entropy}

\vspace{1cm}

\author{\sc Arundhati Dasgupta and Hugh Thomas}
\affiliation{
  Department of Mathematics and Statistics, University of New Brunswick, Fredericton, Canada E3B 5A3}
\email{dasgupta@math.unb.ca,hugh@math.unb.ca}
\begin{abstract} 
We calculate the exact degeneracy of states corresponding to the area operator 
in the framework of semiclassical loop quantum gravity, using techniques of combinatorial theory. 
The degeneracy counting is used to find entropy of apparent horizons derived from generalised coherent states
which include a sum over graphs. The correction to the entropy is determined as exponentially
decreasing in area.  

\end{abstract}
\maketitle
\section{Introduction}
One of the most enigmatic and interesting physical quantities is the entropy of a horizon. As evident now,
the horizons may or may not enclose singularities. From the discovery of the irreversibility associated
with horizon area in any classical process, and laws of black hole mechanics \cite{christd,bek,hawk}, entropy
of a horizon has become synonymous with thermodynamic entropy. This identification requires the search
for a theory of quantum gravity and a corresponding `microscopic' counting of entropy. In the absence
of a complete theory of quantum gravity, the horizon area has been arbitrarily quantised in units of
Planck length squared and a degeneracy of 2 assigned to every bit of area \cite{bek}. Later, reduced
phase space quantisations actually achieved an `area quantisation' \cite{reduced}, and the area spectrum
was identified as equispaced numbers, in terms of the fundamental Planck length squared. However, these
quantisations were not complete `quantum theories' and inclusion of the
full degrees of freedom of gravity might have completely different conclusions from reduced phase 
space quantisations.
 Thus, the search for
an explanation for the microscopic origin of what is now known as Bekenstein-Hawking entropy, which
precisely equals $A_{H}/4 l_p^2$ continues. 
  This search for the origin of entropy is quite obviously linked to the search for a complete theory of quantum
gravity. Within the realities of the four-dimensional world, the most promising theories are the standard
techniques of path-integral quantisation and canonical quantisation of gravity. String theory, the more
radical approach, remains in the realm of the unreal, far from a quantum description of a 4-dimensional
non-supersymmetric, non-charged black hole \cite{stromvaf}. In the path-integral approach, due to difficulties 
in actual evaluation of the functional integral, one has to discretise the space-time. This discretisation of 
space-time for a Schwarzschild
black hole in the dynamical triangulation framework appears in a recent work \cite{bloll}. 
However, it is only in the framework of canonical quantisation of gravity that a derivation of entropy, 
including the resolution of singularity at the center of black hole, has been obtained, mainly prompted by the discovery of
a quantisation of area. For
reduced phase space quantisations using geometrodynamics, there has been recent progress 
in \cite{huwi}. For reduced phase space resolution of black hole singularities in loop quantum
gravity, see \cite{abmo}. However, as shown in \cite{thiembrun}, inclusion of full degrees of freedom
of quantum gravity might change the conclusions completely. Thus, quantisation of horizons is still
a subject of research. Since we are working in the loop quantum gravity framework, much of the
formalism is motivated by a previous derivation of entropy by counting the states of a Chern Simons theory
induced on the horizon boundary \cite{rov,abck}. However, the entire programme as described 
in \cite{abck} of entropy
is characterised by the following\\
a)The microstates for entropy are associated with only the horizon, and the rest of the space-time does not manifest
itself in any way.\\ 
b)In actual black hole processes the horizon is not a boundary of space-time, and information is known to
flow across the horizon in a quantum mechanical description.\\
c)The area eigenstates at the horizon are taken as exact eigenstates, which is a good approximation in the
large area, limit, but should not be the correct physics, if the horizon exists only classicaly.\\
d)The actual counting after the imposition of the horizon boundary conditions on the state were not exact,
and the Immirzi parameter remains a mystery \cite{immirzi,meiss}.\\

However, the calculation was remarkable in the sense that the quanta of area of the horizon was given 
by $4\pi \sqrt{j(j+1)}l_p^2$ with the $j$ corresponding to a spin degree of freedom. For $j=1/2$ this would 
precisely
correspond to the 2 degrees of freedom associated with unit area of $l_p^2$. This would make the previous derivations by physicists a reality. The calculation also showed the way for 
later calculations of black hole entropy in the loop quantum gravity framework. 

One of the recent developments which has a very
correct perspective is to represent an entire black hole space-time by a semi-classical or coherent state \cite{hall,coherent}. 
Since black holes are essentially classical solutions of Einstein's equations, and the light cone
is a classical concept, horizon entropy should be associated with semi-classical or coherent states.
The coherent state is defined as a function of complexified phase space variables, and the expectation
value of operators are closest to their classical value in this state \cite{adg1,adg2}. The horizon is realised as a 
solution
to a difference equation, which is exact only in the classical limit \cite{adg3}. The entire wavefunction for the 
coherent state is defined for a spatial slice of the black hole which includes the horizon as well as the
singularity. The singularity is resolved in the coherent state, and is recovered only in the classical
limit. Most importantly, the entropy can be calculated from first principles as $S_{BH}= - {\rm Tr\rho \ln {\rho}}$,
where $S_{BH}$ is the entropy associated with the horizon, and $\rho$ is the density matrix. This density 
matrix was derived by tracing over the coherent state wave-function inside the horizon. Thus the origin of 
entropy is similar to an entanglement entropy calculation, first observed in \cite{sredn}. Thus it is not 
surprising that the entropy is proportional to the {\it area} of the surface which bounds the set of states 
which are integrated out. What is completely new here is that this entanglement entropy is completely 
`gravitational' in origin and hence finite.
Thus previously infinite calculations of entanglement entropy for scalar fields in black hole space-times 
\cite{ent} are now completely finite and correct. 

However, the coherent state discussed in \cite{adg1} calculated the entropy for one particular graph.
Here, we discuss the situation where, within certain symmetry restrictions, there is a sum over
inequivalent graphs. The kinematic Hilbert space is a tensor sum of individual
Hilbert spaces \cite{astmurthiem}. Each sub-Hilbert space has a `coherent state' corresponding
to that graph. This interpretation is necessary, as within the formalism of \cite{coherent}, the
coherent state is defined over graph dependent variables. The `generalised' coherent state
is defined in the tensor sum of Hilbert spaces, and the operators are also tensor sum
of operators for each graph. For the semi-classical limit, one has to do a suitable
averaging process.  

To obtain entropy, the density matrix is obtained for each graph coherent state separately. 
The density matrix is a tensor sum over density matrices for each graph. 
In the final expression, the entropy is the logarithm of the degeneracy counting associated with the
degeneracy of the area operator at the horizon. This counting is done using generating function 
techniques of combinatorics. The final answers are remarkably simple, and exact. The main observations 
of our paper are\\
i)Entropy is proportional to area of the horizon.\\
ii)The final value of the Immirzi parameter is given as $ 8\ln(3/2 + \sqrt{5}/2)= 8 \ln (2.618)$.
This value is the same value as was obtained in \cite{alek}, where the equispaced area spectrum used here
was used to find the entropy. The ensemble of spins was taken to be made up of completely distinguishable
elements.\\
iii)The corrections to horizon entropy are actually exponentially decreasing with area, in contrast to previously
observed logarithm of area corrections \cite{corr}. The log area correction appears only if the cardinal number of edges inducing
the horizon area are taken to be fixed. However ab-initio, there is no reason
to restrict the number of spins to a particular value. 

Having said all these, one must emphasise that the entropy derivation is done using coherent states
defined on the kinematical Hilbert space, as opposed to the physical Hilbert space where the diffeomorphism
and the Hamiltonian constraints are satisfied \cite{thiem, ditt}. This would be a very interesting formalism
to develop. What we hope is that the generating function techniques of combinatorics used here to calculate the
degeneracy of states and the `semi-classical' realisation of entropy using coherent states will be useful 
for future calculations of entropy. The derivation of the exponentially decreasing correction terms to entropy
is a very new result, and it will be very interesting to determine if this is the theoretical constraint which
will get the correct theory of quantum gravity.

In the next section we give the formalism for the derivation of generalised coherent states, and the sum 
over Hilbert spaces interpretation of the classical 
limit. The third section contains the derivation of entropy from a exact combinatorial counting, 
and finally the fourth section ends with conclusions.

\section{Generalised Coherent states}
In the new-variables or Sen-Ashtekar-Barbero-Immirzi variables, the gravitational canonical variables
of the spatial three metric and the corresponding momentum (function of extrinsic curvature) are 
re-written in terms of densitised triads and a SU(2) gauge connection. The defining
equation for these variables are
\be
\beta E^a_I E^b_J \d^{IJ} = {\rm det q} \  q^{ab} \ \ \ \ A_a^I = \Gamma_a^I - \beta K_a^I
\ee
where $E^a_I$ are the densitised triads, $q_{ab}$ the three metric, $A_a^I$ the SU(2) connection.
$\Gamma_a^I$ is the spin connection corresponding to the triads, and 
$K_a^I= \frac{1}{\sqrt {\rm det q}} e^{b I} K_{ab}$
has the information about the extrinsic curvature $K_{ab}$ associated to the spatial slicing.  
$\beta$ is the Immirzi parameter,
and represents a one parameter ambiguity in the definition of the variables. This ambiguity creates
inequivalent quantisations, and hence the value determined with the black hole entropy is 
in some sense a experimental result for the theory.
The actual quantisation is done in the variables of holonomy and momentum defined on graphs comprised
of piecewise analytic edges, and their corresponding dual graph, comprised of 2-dimensional surfaces.
The basic phase space variables are now:
\be
h_e(A) = {\cal P}\exp (\int A.dx) \ \ \ \ \  P^I_e= \frac1a{\rm Tr}[ T^I h_e\int_S h_{\rho} E h_{\rho}^{-1} h_{e}^{-1}]
\ee
where $h_e(A)$ denotes the holonomy along one edge $e$ and is a path ordered exponential of the gauge
connection along the edge. $P^I_e$ is the corresponding momentum evaluated
on a dual graph comprised of 2-surfaces and is a function of the densitised triads as well as the holonomy
of the gauge connection along the edge $e$ which cuts the dual surface, and the $h_{\rho}$
denote the holonomy along the edges defined on the dual 2-surface S. $a$ is a dimensional constant,
which is determined from the length scales of the theory.

Thus in the quantisation of the above variables, `graphs' have a very important role. 
In generic
quantisation, the Hilbert space is taken as defined over the configuration space, which is the
projective limit of graphs. This is possible as the graphs form a directed set $\g'> \g$ where
$\g'$ can be obtained by sub-dividing the edges of $\g$. The projection operators $p_{\g'\g}$ then
map the graph configuration spaces. The kinematic Hilbert space is defined over the space of the
connections modulo the gauge transformations. The states in this space are not diffeomorphism
invariant, and neither do they satisfy the Hamiltonian constraint. However, they are relevant,
as they have a correct semi-classical limit, and the expectation value of operators are closest
to their classical values, evaluated on a classical metric.

Now, these variables have a Poisson algebra which can be lifted to commutators as a quantisation. The 
quantised Hilbert space is given by orthonormal basis states $|jmn>$ for each edge. A complexification
of the phase space takes the variables to SL(2,C) valued elements 
\be
g_e = \exp^{-i T^I P^I/2} h_e
\label{eq:clas}
\ee
($T^I$ are the generators of SU(2) and are $\frac{-i\sigma}{2}$ where $\sigma$ are Pauli matrices) and the Hilbert space wavefunction can 
be defined in terms of these in the Segal-Bergmann representation. 
The kernel of
the transformation, which is also called a coherent state transform, is the coherent state itself, as
originally defined by Hall. This of course is a function of both $h_e, g_e$, where now $g_e$ is interpreted
as a classical label for the phase space point at which the wavefunction is peaked. Now, given a graph
$\G$ comprised of edges, the complete coherent state for the entire graph is given by a tensor product 
over coherent states for each edge. Thus,
\be
\Psi^t(g_e,h_e) = \prod_e \psi^t(g_e,h_e)
\ee
($t=l_p^2/a$ is the semiclassical parameter.)
Imposing the trivial representation of intertwiners at the vertices would give a gauge invariant 
state, and the expectation values of the operators can be determined using the same techniques
as above, and they are also closest to their classical values \cite{thiemwink}.
The coherent states written down in \cite{adg1,adg2,adg3}, were defined for a Schwarzschild black hole,
also in a particular choice of coordinates, and one particular choice of graphs. The entropy was also
defined for one particular graph. 

However, there is no reason to choose one particular graph: 
the graphs can be varied. Two graphs, which belong to a
directed set, can be said to be equivalent, as the expectation values of the holonomy and the momenta can be 
obtained by taking the product of the holonomy and the momenta of the other graph. In a directed set
$\g'> \g$, the edges of $\g'$ are obtained by subdividing the edges of $\g$. Since, for a edge $e$, which 
is 
broken into two edges $e_1, e_2$, the holonomy would be $h_e= h_{e_1} h_{e_2}$, one can always obtain 
operators of $\g$, if one knows the operators of $\g'$. However, the reverse is not true, and hence, given
a projective family one can always identify a `minimal graph'. The kinematic Hilbert space is then a tensor
sum of Hilbert spaces for these minimal graphs. 

Each such Hilbert space can be characterised by a coherent state, and a corresponding classical limit obtained 
for the operators as defined in terms of the expectation values in those states. The generalised coherent state 
would be defined on the tensor sum of Hilbert spaces.
However, there is a subtlety about the graphs and the classical limit of operators defined on them, which 
first has to be completely understood. 
The graphs in consideration for the classical limit have to be regarded in terms of two separate questions.\\
1)The actual choice of a `minimal' arbitrary graph might be classified as follows.\\
i)The number of vertices.\\
ii)The number of edges linked at a particular vertex, or the valency of the graph.\\
iii)The orientation of the edges.\\
iv)The total number of edges, which can in principle be countably infinite.\\
v)Any symmetries of the graph.
An example of this would be a cubic graph, which is six valent, and can be very useful in determining
3-dimensional semi-classical limits, as considered in \cite{hanno}.\\
2)The actual embedding of the graph in a classical metric. These would be decided by the following considerations:\\
i)The classical metric.\\
ii)The edge lengths of the graph as measured in the classical metric.\\
iii)Symmetries of the classical metric.\\
iv)The holonomy and the momentum, which will characterise the physical phase space.\\

For example, a cubic graph can have a spherical embedding, as well as a non-spherical embedding.
Thus, there is no unique way of embedding a given graph in a classical space time, and hence the same
`quantum graph' can correspond to different `classical graphs'. In the coherent states for gravity, the
actual classical graph can be thought as labelled by the $g_e$ (\ref{eq:clas}). Thus, the semi-classical 
limit
corresponding to a given Hilbert space will be characterised by the graph, as well as the set of labels
denoted as $\{g_e\}$.

The graph as discussed in \cite{adg1} for spherically symmetric classical space-times 
was actually fixed by the symmetries of the classical metric. The 
graph was chosen so that the $n$th vertex coincided with the 0th vertex. Further,
each vertex was 6-valent. 
While taking the `classical limit' of this graph, what remained fixed was the cardinal number of edges, and
vertices, and the above stated `spherical symmetry'.
Thus, each coherent state, corresponding to a given graph, will now be characterised by
\be
\Psi_{\G,\{g_e\}},
\ee
where $\G$ denotes a minimal graph.

In case of the spherically symmetric embedding, the choice of edges along the coordinate lines of
$r,\th,\phi$  of the sphere ensures that the graph samples maximal points near the origin. However, here, 
the
number of vertices is fixed per volume, and hence the density decreases as one approaches the
asymptotics. To sample more points in the exterior region, new non-radial edges have to be introduced.
The questions of singularity resolution and entropy calculation are however quite relevant as given in
\cite{adg3}. Also, a very similar choice of graph seems to be sufficient to describe the interior of
a reduced Schwarzschild space-time in \cite{abmo}.
Thus, even with a rather careful choice of graph, there remains the ambiguity of the classical graph.

The question is, given a graph $\G$, how does one evaluate the sum over classical labels $g_e$?
For coherent states, there is a very natural measure for the classical labels, given by the Liouville
measure, and this was discussed in \cite{adg3}. The answer obtained there is that the area induced
by the edges given by $\sqrt{P^IP^I}=P_e$ has to be very small. This can be interpreted as an indication
of a natural way to ensure microscopic sampling. However, in a generic situation, a graph can induce
all possible values of $g_e$. In the very particular restrictive case we consider here, we give each
graph a unique classical set of labels $\{g_e\}$. If we take the Liouville measure seriously, then,
as considered in \cite{adg3}, only those graphs which induce the minimum area of the 2-surfaces 
would be preferred. In the coherent states area spectrum, this would correspond to $j=0$ or
semiclassical area $t$. This would not yield any entropy associated with the horizon. However, 
 given any graph, we choose the classical graph to sample the 
minimal areas induced.

Thus a generalised `coherent state' will be:
\be
\Psi= \otimes_{\G,g_e} \Psi_{\G, g_e}
\label{eq:prod}
\ee
The symbol $\otimes$ denotes cartesian product, and the operators will be tensor sums. The eigenvalues
will then be just sums of eigenvalues for each space.
Now, we concentrate on the `annihilation operators' for such Hilbert spaces, and determine their
action on the coherent states. This annihilation operator is defined as 

\be
\hat g_e= e^{3t/8} e^{-i \hat P^j_e T_j/2} \hat h_e
\ee

In the tensor sum Hilbert space, this would correspond to having only the block belonging to this particular 
graph edge as operator valued in the tensor summed operator.
Hence, this operator of course is specific to a particular edge, and hence would give a non-zero value, iff a particular
graph containes the edge. The local properties would be very specific to the particular graph, and hence 
the edges.

\be
\hat g_e |\Psi>= g_e |\Psi>
\ee

Thus, only the particular graph edge corresponding to that particular operator will give a non-zero eigenvalue.
The variable $g_e$ can then be used to estimate the values of the holonomy and the momenta using 
the following:

\be
e^{-i T^I P^I}= g_e g^{\dag}_e e^{-3 t/4}
\ee
and
\be
h_e= e^{-3t/8}e^{i P^I T^I/2} g_e
\ee

Now in the generalised coherent state, one has the opportunity of improving the estimate obtained from a single graph, which might not sample
all possible points of the classical metric. If one takes the annihilation operator which is the tensor sum
of annihilation operators of all the sub-Hilbert spaces, then in a given region $R$, at least one of the edges
will sample the points. Thus for the `global' annihilation operator, which will be a tensor product of 
annihilation operators for edges of each individual graph, there will be a maximal sampling of the 
classical points in the tensor summed Hilbert space.
 
There will, however, be a overcounting for `macroscopic' operators, whenever the same edge samples the same classical points in more than one graph. 

This
will be evident, in the `global operator' or the tensor product of the operators for each complete graph.

Let us explain what we mean by a `global operator'.
The annihilation operator corresponding to a entire graph would be
\be
\hat g_{\G}= \otimes \hat g_{e}.
\ee
This operator will have eigenvalues which are 2 $\times$ 2 matrices for each edge. Thus if the
graph has $n$ edges, the tensor product will be a $2n \times 2n$ matrix.

If there corresponds another minimal graph, which has $m$ edges, the corresponding `global operator' would be
a  $2m \times 2m$ matrix.

The tensor sum of the operators would be to simply add the $2n \times 2n$ and the $2m \times 2m$
matrices to obtain:

\be
 \hat g_{\G, 2n} \oplus \hat g_{\G,2m}.
\ee

Now, what is mainly interesting is the situation where the tensor sum of operators is being used to 
measure classical quantities. If the two graphs have very similar embeddings in the classical metric,
then since many of the edges sample common classical points, the eigenvalues will be repeated. For measurement
of macroscopic quantities, e.g. area of a 2-sphere of radius $R$, both the graphs would give the same 
result to first order in the semi-classical parameter.

In the situation considered in \cite{adg3}, this can be the area of the horizon. Each of the graphs will
give the area as $\sum_e P_e=A$, and they should all be equal. Thus, though the number of edges will differ
from graph to graph, due to the macroscopic area restriction, the total area has to be same. Thus
for the total area operator defined for the 2 graphs above, in the limit $t\rightarrow 0$,
\be
<\Psi| \hat A \oplus \hat A|\psi> = 2 A
\label{avg}
\ee
Thus, the operator has to be divided by the number of graphs in the tensor sum to obtain the correct value. 
The point we are trying to make is that the local properties can be recovered from any set of edges, and
their expectation value on a coherent state adapted to that graph. 
However for operators, charting macroscopic or global regions of the manifold, the tensor sum of operators
will sample the same classical region more than once, and hence one has to find a suitable averaging
process.
Now, in principle, there can be
countably infinite graphs, and hence the averaging procedure could be a `infinite renormalisation'
process, which physicists are quite familiar with. However, as we discuss, the symmetry restrictions
of the classical metric might restrict the number of graphs, and thus there will be a method
to count the graphs.

Now we specialise to the case of the spherically symmetric apparent horizon considered in \cite{adg3},
and observe what a generalised coherent state entropy would correspond to.

 The horizon area is induced
by radial edges crossing the horizon, as was motivated from the apparent horizon equation in \cite{adg3}. 
This is really a very restrictive example of inducing area of one particular surface by a set of edges. 
The number 
of degrees of freedom which can vary is really the number of radial edges, and their spacing on the surface.

Let us represent that by a planar circle, and see how the graphs can vary. 
There are actually different ways to change the graph. These can be of the following form\\
(i) The cardinal number of edges changes. Here, as explained in the following
diagram, the introduction of two new radial edges (in bold and in color) 
crossing the horizon,
is actually changing the previous minimal graph to another one, as clearly
the new graph is not obtained by subdividing previous edges. Hence the
two Hilbert spaces corresponding to the graphs would have to be summed over.\\

\includegraphics[scale=0.7]{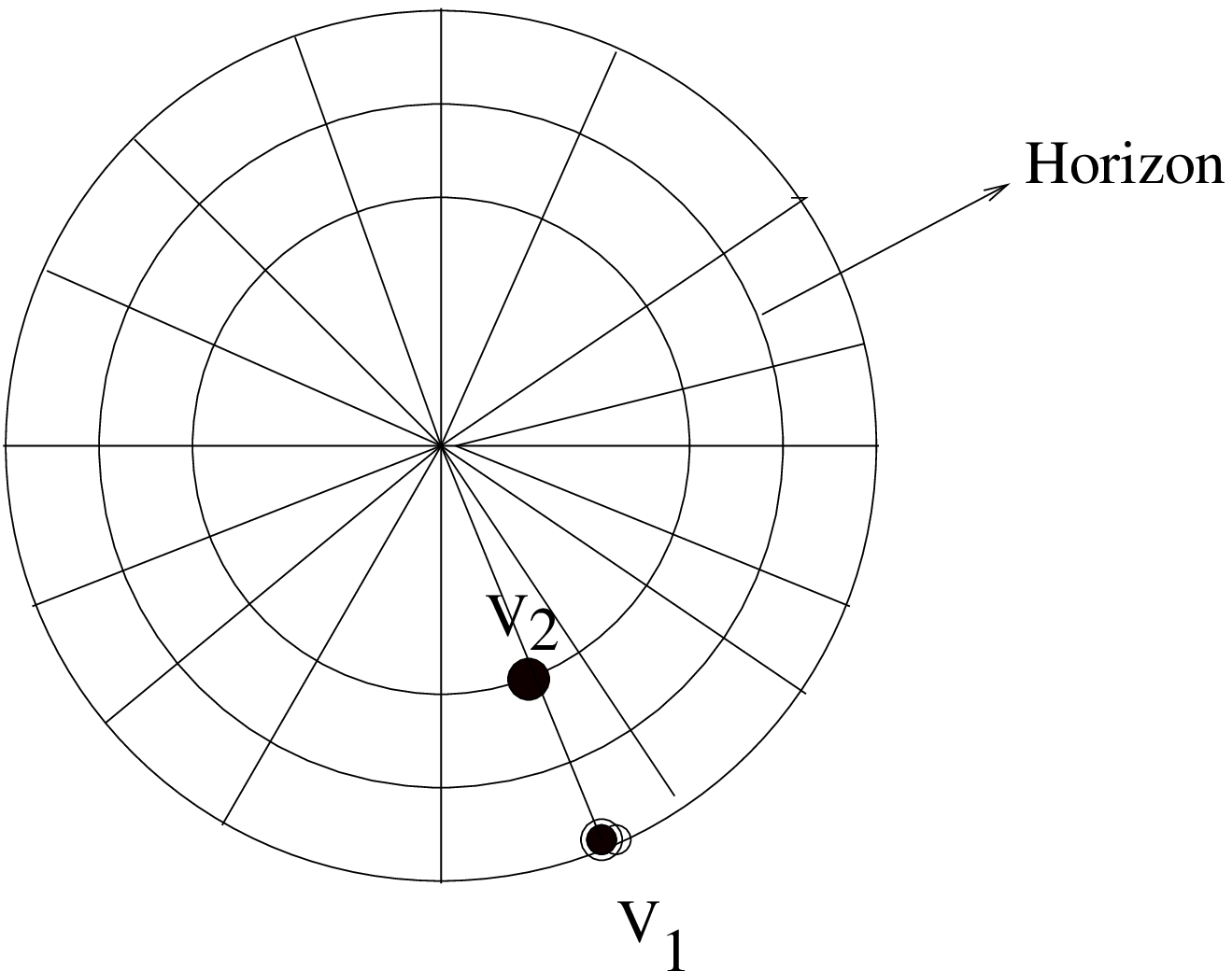}

(ii)The graphs have the same number of edges, however these are not
symmetrically embedded. Hence the spin assignments $j_{cl}$ will differ.
In other words the classical labels $g_e$ would be different. This would 
however still be a `different' minimal graph, which we explain subsequently.

\includegraphics[scale=0.6]{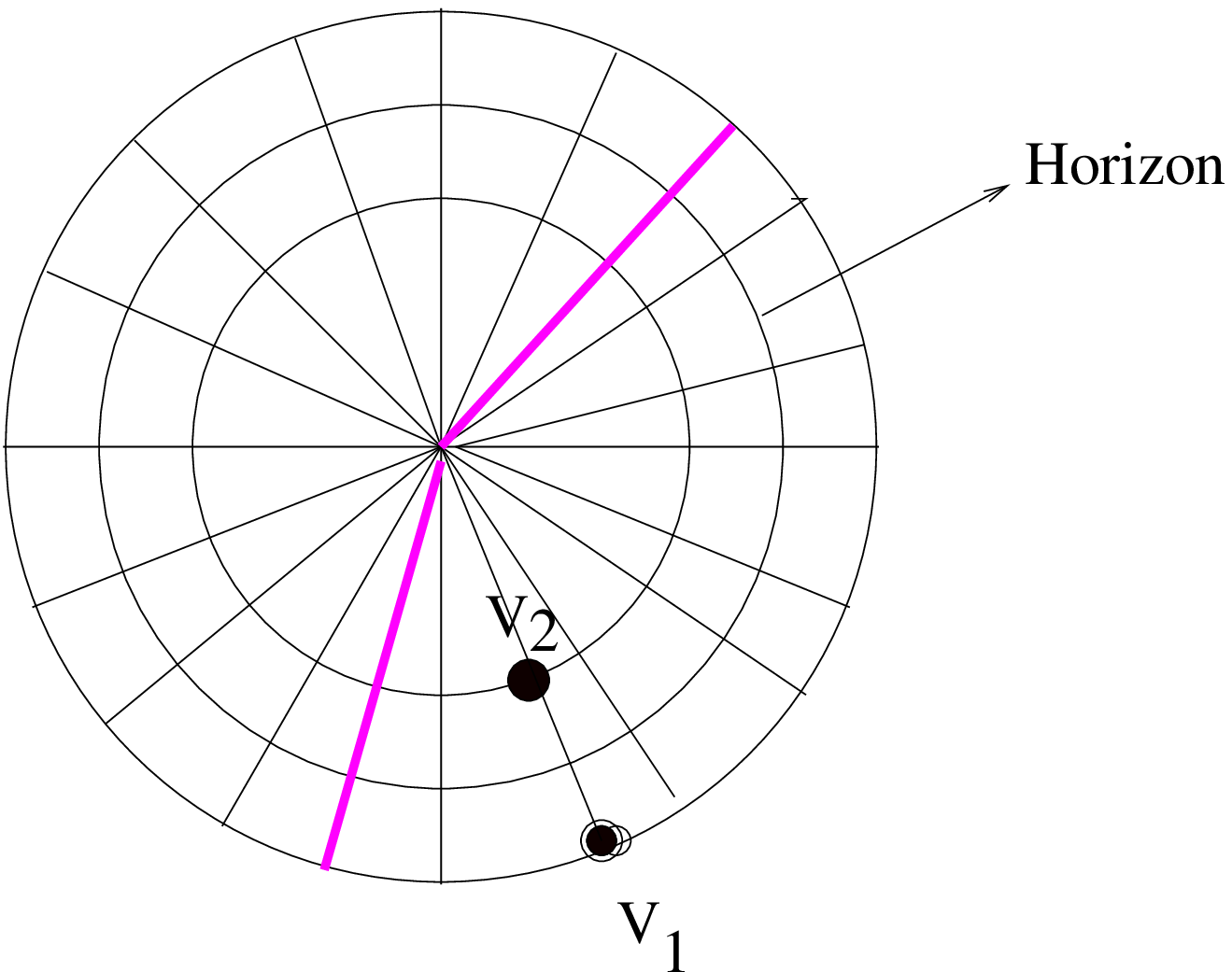}

The reason we are concentrating on the graph at the horizon is because 
entropy is only a function of the edges at the horizon, and these are a countable set. Hence all graphs
which have the same number of edges with the same set of classical labels $g_e$ at the horizon, can be grouped
together as entropically `equivalent class' of graphs and these will be quite definitely countable. Hence for the purposes of the calculation of black hole
entropy, we will sum over only those graphs which differ at the horizon.

To understand the tensor product structure using a scheme given in \cite{ashtlewand}, we find that
the Hilbert space can be written as a tensor sum of Hilbert spaces for each graph $\alpha$, 
with a restriction that $H'_{\alpha}$ consists of cylindrical functions, which are orthogonal to 
every other graph strictly contained in $\alpha$. Thus $H'_{\alpha}$ is a tensor sum over states with non-zero
$j$ for each edge, and non-zero representation $l'$ at each vertex. Now, let us look at the following graphs
at the horizon, and identify the `minimal graphs'. Let us start with a graph with $n$ radial edges crossing the horizon.
If one tries to reduce the number of edges crossing the horizon, since the embedded graph is six-valent,
one has to remove an entire set of vertices along one radial coordinate. 
 If the graph has different valences away from the horizon, 
removal of vertices immediately outside the horizon, and the vertices within would
have the same effect. Now, let us start from a graph with $n$ edges, and with a set of removal
or addition of vertices, see how different graphs can be obtained from the same. 
For example, removal of the $i$th edge, $i-1$th edge, $i-2$th edge gives a graph of $n-3$  
edges, however, this graph
would be inequivalent to the graph obtained by removing the $i$th edge, $i-4$th edge and $i-8$th edge from 
the
same graph. Graph I would differ from Graph II in the distribution of the areas induced by the edges at the
horizon 2-surface in the classical limit. One thing is very clear, since the edges are embedded in the classical metric, the areas
induced by the edges would correspond to `distinguishable' spins. The Hilbert space of graph I would have the
same number of edges as graph II crossing the horizon, however, they are inequivalent graphs and their 
semi-classical limits are different.  

\includegraphics[scale=0.5]{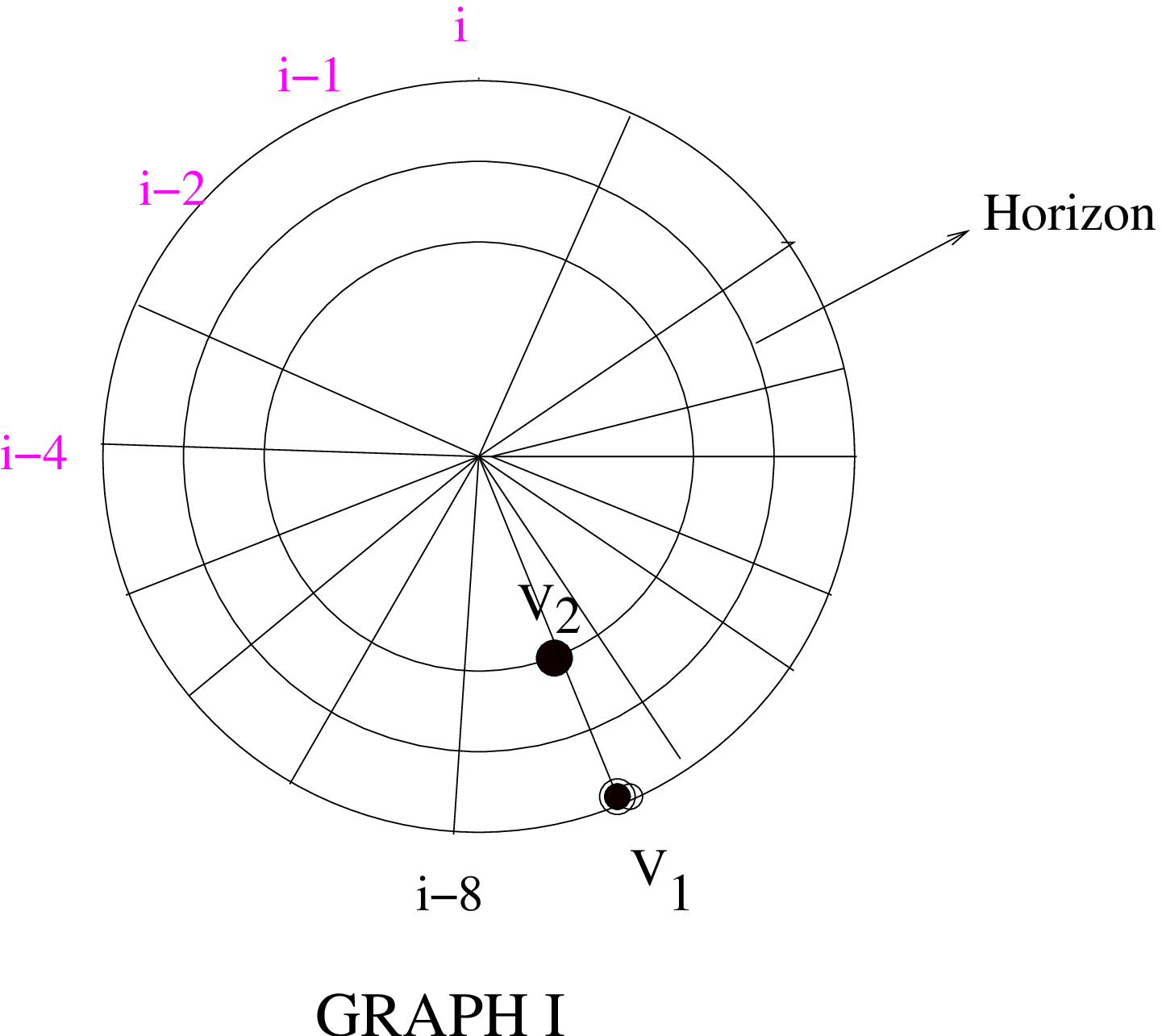}\\
\includegraphics[scale=0.5]{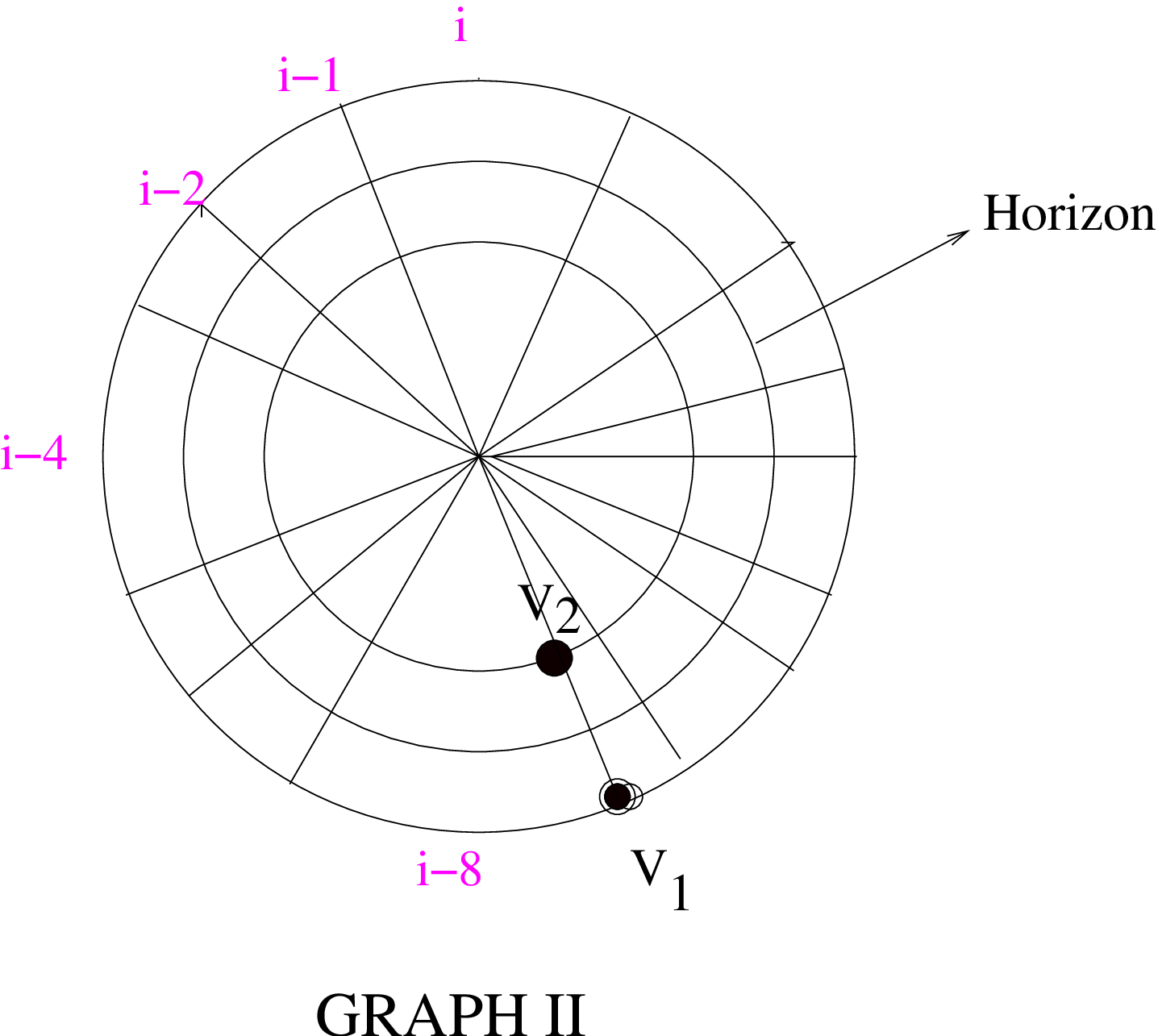}

Now, one would think that the density matrix will be a tensor sum over the density matrices for each graph. 
The tracing
process of edges `inside' the horizon is graph specific. Hence, each Hilbert space will be associated 
with its own density matrix, and the combined density matrix will be a tensor sum of the density matrices
for each graph
\be
\rho= \oplus \rho_{\G_{n,\{g_e\}}}
\label{ge}
\ee
Now, thinking of this as derived from a superposition of coherent states for different graphs, this
is equivalent to ignoring interference effects between two different graph Hilbert spaces. We 
calculate the entropy corresponding to the above, and find the leading term to be Bekenstein-Hawking. 
The corrections are exponentially decreasing in area. However, if one includes 
those, one gets a larger entropy. The leading term is again Bekenstein-Hawking, and the corrections
are exponentially decreasing in area. We now give a much more detailed description of the derivation
of entropy.

\section{Entropy of Horizons}
The entropy associated with spherically symmetric apparent horizons has been derived in \cite{adg3}.
The apparent horizon equation shows that the `classical correlations' across the horizon have information
only about the total areas induced at the horizon by the edges. The individual components $P_H^I$ remain
undetermined. This is the origin of degeneracy in the density matrix. Each individual Hilbert space
has its own density matrix, however, the total entropy should include a sum over all possible graphs.
As obtained in \cite{adg3}, the density matrix for each graph is of the form:
\be
\rho_{\G,\{g_e\}} = \prod_{i=1..n}\rho_e
\label{comp}
\ee
where the edges are precisely the edges crossing the horizon and inducing the same with tiny
bits of area, given in terms of the spin as $P_H= (j_{\rm cl} + \frac12) t$ where $t= l_p^2/a$.
Each $\rho_e$ is diagonal, with the entries being $1/(2 j_{\rm cl} + 1)$. 
We give the derivation here, for the density matrix for one particular graph, which has say $n$ 
radial edges crossing the horizon. Ab initio, the tensor product of the coherent states
peaked at the individual edges does not have information about the continuity of the classical
metric across the vertices. This has to be included in the definition of the semi-classical limit
of the quantum theory, as the classical metric is derived from a local differential equation.
Hence, the metric of two neighbouring points $x^{\mu}$ and $x^{\mu} + \epsilon$ are obtained by
solving the equation of motion, or the differential equation corresponding to the Einstein Lagrangian.
Thus the holonomy of edge $e_1$ is related to the holonomy of edge $e_2$, if they are linked at the
same vertex. However, as in \cite{adg3}, we are interested in determining the correlations in the 
wavefunction which exist due to the existence of an apparent horizon equation. The classical metric
beyond the horizon, and before the horizon, do not contribute in any way to the correlations across
the horizon. Since the trace is taken over the wavefunction inside
the horizon, the correlations which matter are the ones which link the wavefunction of the edges
outside the horizon with those immediately inside the horizon. This is simply obtained by identifying
the apparent horizon in the spherically symmetric coordinates, and then identifying the correlations
by lifting the equation to an operator equation in $h_e, P^I_e$, which is also a discretised version 
of the same. Thus the classical metric of space-time is chosen in a particular set of coordinates
where the spatial or constant time slices are spherically symmetric and have 0 intrinsic curvature:
\be
ds_3^2= dr^2 + r^2 (d\theta^2 + \sin^2\theta d\phi^2)
\ee
The entire curvature of the space-time metric is contained in the extrinsic curvature or $K_{ab}$ tensor.
Now, if there exists an apparent horizon somewhere in the above spatial slice, then that is located as
a solution to the equation
\be
\nabla_a S^a + K_{ab}S^aS^b -K=0
\ee
where $S^a$ , ($(a,b=1,2,3)$ denote the spatial indices) is the normal to the horizon. If the horizon is chosen to be the 2-sphere, then $S^a\equiv(1,0,0)$,
and the apparent horizon equation as a function of the metric reduces to:
\be
K_{rr}(1-q^{rr}) -K_{\phi\phi}q^{\phi \phi} - \G^{\phi}_{\phi r} -K_{\th\th}q^{\th \th}-\G^{\th}_{\th r}=0
\ee
The first term of the equation disappears trivially as $1=q^{rr}$ for any point in the spatial slice.
Hence the extrinsic curvature along the radial direction plays absolutely no role in the apparent horizon
equation. In the discrete version of the above equation, the information about the extrinsic curvature
is contained in the holonomy along the radial edge. The $h_{e_r}$ play absolutely no role in the
correlations induced in the coherent state wavefunction due to the apparent horizon equation.
The radial wavefunction is thus independent and is not restricted by the apparent horizon equation.
The rest of the equation is converted to a difference equation in the variables $h_{e_\th}$ and $h_{e_\phi}$,
$P_{e_\th}$ and $P_{e_\phi}$ by a suitable regularisation along edges. This equation, as described 
in detail in \cite{adg3}, is given by
\bea
&& \frac{P_{e_\th}^I P_{e_\th}^I}{V}\left[Tr\left[ T^J h_{e_\th}^{-1}[h_{e_\th}, V]\right]_{v_1} - Tr\left[T^J h_{e_\th}^{-1} [h_{e_\th},V]\right]_{v_2}\right]Tr[ T^J h_{e_\th}^{-1}[h_{e_\th},V]]_{v_1} \nn \\
& & \ \ \ \ \ -\frac{1}{\sqrt \beta}\frac{\partial}{\partial {\beta}} Tr[T^I \  ^{\beta} h_{\th}] \ ^{\beta}P_{e_\th}^I + (\th \rightarrow \phi)= 0 \nn \\
&{\rm or}& 4 P_{e_\th}^I P_{e_\th}^I \left[Tr\left[ T^J h_{e_\th}^{-1}[h_{e_\th}, V^{1/2}]\right]_{v_1} - Tr\left[ T^J h_{e_\th}^{-1}[h_{\th}, V^{1/2}\right]_{v_2}\right] Tr[T^J h_{e_\th}[h_{e_\th}, V^{1/2}]]_{v_1} \nn \\
&& \ \ \ \ \ \ -\frac{1}{\sqrt \beta}\frac{\partial}{\partial \b} Tr[T^I \ ^{\b} h_{e_\th}] \ ^{\b} P_{e_\th }^I + (\th\rightarrow \phi) =0
\label{diff}
\eea 
where quantities evaluated at a vertex $v_1$ outside the horizon get related to quantities inside the
horizon at a vertex $v_2$. The appropriate regularisations which have been used here are given by
\be
K_{\th(\phi),\th(\phi)}= e^I_{\th(\phi)}K^I_{\th(\phi)}, \ \  q_{\th(\phi)\th(\phi)}=e^I_{\th(\phi)}e^I_{\th(\phi)}
\ee

\be
e^I_{\th,(\phi)}= Tr[T^I h^{-1}_{e_{\th(\phi)}}\{h_{e_{\th(\phi)}},V\}]
\ee

and

\be
K^I_{\th(\phi)}=\beta\frac{\partial}{\partial\beta}Tr[T^I h_{e_\th}]
\ee
It is quite difficult to explicitly solve (\ref{diff}), but one can gather some information from them,
 in the following way. Symbolically (\ref{diff}) can be written as
\bea
A[B^I(v_1)-B^I(v_2)]C^I(v_1)-D &= & 0 \\
B^I(v_1) C^I (v_1)-\frac{D}{A} & = & B^I(v_2)C^I(v_1) 
\label{cor}
\eea
Now, though the equation is completely independent of the direction $I$ in the internal gauge space,
the individual components of $B^I(v_2)$ are dependent on the components of $B^I(v_1)$ as projected
along the vector $C^I$. Thus, the apparent horizon equation solves even the internal gauge components
of $g_e$ or the group valued classical label of the edges ending at vertex $v_2$ in terms of the
classical labels at vertex $v_1$. 

Now, the apparent horizon equation (\ref{diff}) is thus a difference equation relating the canonical variables
of holonomy and momenta at vertices $v_1$ with vertices $v_2$ which are angular in nature. Inspired by
this, and as sketched before, the graph at the horizon is particularly chosen to be comprised of
only {\it radial edges} which cross the horizon, and link vertices of the type $v_1$ with vertices
of the type $v_2$. When the above is lifted to an operator equation, and the expectation value of the
operator equation evaluated in the coherent states, one ends up correlating the wavefunction outside
the horizon with those inside. In the $t\rightarrow 0$ limit, this is manifested as a relation of the
classical labels outside with those inside.
\be
g_{e \ v_2}=g_{e \ v_2}(g_{e \ v_1})
\ee

This correlation is introduced in the coherent state wavefunction by a conditional probability function
$f(g_{e \ v_1},g_{e \ v_2}|P_H)$, which takes care of the classical limit only, and is the same as a number
$N$ in the $t\rightarrow 0$ limit. This is done to retain the form of the coherent state wavefunction
as defined in \cite{thiem}, and still have information about the correlation.
For the calculation of the density matrix, the coherent state can be written as the tensor product of 
Hilbert spaces associated with graph edges which end at a vertex $v_1$ outside the horizon, the edges
which start at a vertex inside the horizon $v_2$, and the radial edges $e_H$ which link the vertices
$v_1$ and $v_2$. The wavefunction, in the momentum representation, can be written in the area eigenbasis
which are $|jmn>$ where $j$ is the SU(2) Casimir quantum number and $m,n =-j...+j$.
\be
|\Psi>= \sum_{\{j_O\},j_H,\{j_I\}}\psi_{\{j_{O}\},j_H,\{j_{I}\}}|\{j_O\}>|\{j_H\}>|\{j_I\}>
\ee
where the $\{j_O\}$, $\{j_I\}$ denote all the combined Casimirs of edges outside and edges inside, and $j_H$ is the
horizon Casimir eigenvalue. Now the density matrix formed from them can be written as
\be
\rho= |\Psi><\Psi|
\ee
If the coefficients $\psi_{\{j_{O}\},j_H,\{j_I\}}$ factorise into products of coefficients, then any reduced
density matrix obtained from them by tracing over one of the Hilbert spaces will be again a pure density matrix.
However, if it does not factorise, then the resultant density matrix is a mixed density matrix. In the
formulation of the coherent state in \cite{thiem}, these would have factorised. However, now, due to the
apparent horizon equation, these do not factorise, and the reduced density matrix is a mixed density matrix.
Now, in the $t\rightarrow 0$ limit, the density matrix simplifies to just a diagonal one, with only non-zero
elements in a $(2 j_{cl}+1)\times (2 j_{cl}+1)$ block, where $j_{cl}$ is the `Casimir' corresponding to the
classical value of the area induced by the radial edge on the horizon. 
In the $t\rightarrow0$ limit, the wavefunction can be written in the form
\be
|\Psi>= \sum_{\{j_O\},\{j_H\},\{j_I\}}\psi_{\{j_O\}}(g_{\{O\}})\psi_{\{j_H\}}(g_H)\psi_{\{j_I\}}(g_I)f_{\{j_O\},\{j_H\},\{j_I\}}(g_o,P_H,g_I)|\{j_O\}>|\{j_H\}>|\{j_I\}>
\ee
Where the wavefunction is in a semifactorised form, with the information about the correlation only in the
term $f$, which prevents a complete factorisation of the coefficients. If the $g_I$, $g_O$ is any different
from that determined by the apparent horizon equation of (\ref{diff}), then the function is vanishing, otherwise
it is non-vanishing. Now, the main point to note here is that since the apparent horizon equation is independent
of the holonomy of the edges at the horizon, and depends only on the eigenvalue $P_H$, the $f$ function does
not have any information of the eigenvalues $m_H,n_H$ associated with the horizon edges. However, this is clearly
not true for internal quantum numbers $m_{\{O\}}, n_{\{O\}}, m_{\{I\}},n_{\{I\}}$ as these determine the 
internal component along the gauge directions, and hence by (\ref{cor}) have to be determined. Thus the
degeneracy of the density matrix in the $t\rightarrow0$ limit arises 
only from the horizon edge quantum numbers.

Now, using the explicit form of the coefficients $\psi_{jmn}=e^{-t j(j+1)/2}\pi_j(g)_{mn}$, and the fact that
these are non-zero in the classical limit only for certain values of $\{j_O\},{\{j_H\}},{\{j_I\}}$, the
density matrix has diagonal terms only for
\be
\rho_{\{j_O^{cl}\},j_H^{cl},m_H,n_H, \{j_I^{cl}\}}= |f_{\{j_O^{cl}\},j_H^{cl},m_H,n_H, \{j^{cl}_I\}}|^2
\ee
Here the $m_H,n_H=-j^{cl}_H...j^{cl}_H$. Thus, to obtain $Tr\rho=1$, $|f|^2=\frac{1}{2j^{cl}_{H}+1}$. This completes
the simple derivation. Thus, in the end, the entropy derived from the above is the number of ways to
induce the horizon area, in the semi-classical limit. One has just given a framework for obtaining the
same, using a coherent state wavefunction, and a method of tracing the wavefunction within the horizon.
The points to note here are, of course, the coherent states are still kinematical states, and the
Hamiltonian constraints have not been imposed. This work still requires much more effort, and work is in
progress. The main observation here is that entanglement entropy can be completely gravitational
in origin, and when so, can be completely finite.

Now the above is the density matrix obtained locally, or due to the tracing over the wavefunction for one
particular radial edge. The complete density matrix would be given by (\ref{comp}). 

The entropy function for the density matrix would be
\be
S_{BH} =  - {\rm Tr}(\rho \ln \rho) 
\ee

Now, if the number of spins were $i=1..n$, then the density matrices would be
 $\prod (2j_i+1) \times \prod (2j_i+1)$ (dropping the classical labels) 
matrices with diagonal entries $1/\prod (2 j_i +1)$, and 
hence the entropy would be of the form:
\begin{eqnarray}
S_{BH}&=& \sum_i\ln(2j_i+1) \nn \\
&=& \ln(\prod_i (2j_i+1)) \label{ent1}.
\end{eqnarray}
In addition there will be the constraint that the spins sum up to the total area of the horizon
$\sum_i(j_i+1/2)t= A_H/l_p^2$.
Similarly when the `generalised' density matrix is considered, the density matrix is not obtained
as a tensor sum of the above density matrices for each individual graph. That would be a mixed sum,
as the interference terms of the different Hilbert spaces would not be there. However, we just
consider the same here,

\be
\rho= \oplus \rho_{\G}
\ee
The combined density matrix has block diagonal entries each of dimension $\prod_i (2 j_i +1) \times \prod_i ( 2 j_i +1)$ 
dimensions with diagonal elements of $1/\prod_i(2j_i +1)$. The trace of the density matrix has to be normalised
to be 1. This implies the diagonal elements still have to be divided by $1/d$, where $d$ is the total number
of configurations. The entropy is then
\be
S = - Tr(\rho\ln\rho) = \frac{1}{d}\sum_{j_i}\ln(\prod_i(2 j_i + 1)) + \ln d
\label{deg}
\ee
The leading contribution to entropy appears from the $\ln d$ term, where d is the total number
of configurations. This counting can be exactly done and equals $2^{N-1}$, where $N= 2 A_H/l_p^2$. The other term can be shown to be bounded by a 
constant independant of $A_H$.  
This gives the entropy to be linear in area.  Both of these calculations
are carried out in (\ref{count}) below.

To see what a more `democratic' density matrix might look like, we now
derive the density matrix from an ab-initio `generalised' coherent state, which is a state in the tensor sum of Hilbert
spaces. 

The coherent state for a generic graph is a tensor product of coherent states for each edge comprising the
graph. The `generalised coherent state' is then a tensor sum of the above coherent states.
\be
|\Psi_{\rm gen}> = \oplus_{\G} |\Psi_{\G}>
\ee

The density matrix written for the above would be as given below:
\be
\rho_{\rm gen} =  |\Psi_{\rm gen}><\Psi_{\rm gen}|.
\ee

The above does not break up into a tensor sum of the following
\be
\rho_{\rm gen} = \oplus_{\G}|\Psi_{\G}><\Psi_{\G}|
\ee
but there remain cross terms in the metric, which can give rise to transition amplitudes from one graph to another.
This is difficult to physically justify at the semi-classical level, however, at any finite order in
$t$ where there will be quantum corrections, such transitions are completely possible. A given graph can
have an edge added to it due to an operator acting on it; for example, 
the Hamiltonian operator adds edges to a given
graph. Thus to zeroth order in $t$, the density matrix should be completely diagonal and as $t$ becomes finite,
it 
will have off-diagonal terms also appearing. Now, to calculate the terms, the basic procedure would be the
same as outlined in the previous section, however, now, what will be interesting is the correlations across the
horizon, considering that the different graph sectors can transform into each other. Thus the tracing has to
be carefully implemented, including the correlations across the horizon. Now the density matrix in the previous
case was derived in a local way, with the tensor product structure giving a diagonal matrix in the end.
Here the procedure will be quite similar, except that the correlation function $f$ has to be implemented
in the generalised coherent state.
\be
|\psi_{\rm gen}> = \oplus \prod_{e_{v_1} e_H e_{v_2}} |\psi_{e_{v_1}{e_H}{e_{v_2}}}>
\ee
where the above corresponds to the set of edges at the horizon for each inequivalent graph considered
above.

Now, the correlation function $f(g_{\{O\}},g_{\{I\}}, P_H)$ which is introduced to approximate the 
correlated coherent state to implement the `local apparent horizon equation', now has the added 
complication to ensure that they are independent of the graph $\G$ to which the edge belongs.
As obtained in the previous calculation, if the correlation function at the classical level
depended on the graph, then it would have to include global properties, such as, for example, the total number
of edges crossing the horizon, the distribution of the vertices in the classical metric, etc, which
clearly is not true.
\be
\oplus \prod_{e_{v_1} e_H e_{v_2}} f_{g_{\{O\}},P_H, g_{\{I\}}} \psi_{e_{v_1}}\psi_{e_H}\psi_{e_{v_2}}
\ee
 
In the classical limit, the derivation of the density matrix will follow the exact 
procedure as above, and the off-diagonal elements for different graphs would be damped in the
semi-classical limit. Thus in the $t\rightarrow 0$ limit the density matrix will comprise only of the
diagonal elements, but now, each of them proportional to the same function $|f|^2$. 
The density matrix thus obtained is a $\sum_{\{j_i\}} \prod_i(2 j_i + 1)$ $\times$ 
$\sum_{\{j_i\}} \prod_i (2 j_i +1)$ matrix. The sum is over all possible configurations of the horizon graph.
 The density matrix has diagonal entries, each proportional to
$|f|^2$ hence, 
from ${\rm Tr \rho}=1$, one obtains the individual diagonal entries to be $1/(\sum_{\{j_i\}} \prod_i (2 j_i +1))$,
.
The entropy obtained from this density matrix is then
\be
S = - {\rm Tr}(\rho \ln \rho) = \ln (\sum_{j_i} \prod_i (2 j_i +1))
\label{ent2}
\ee

The combinatorial counting for this is calculated exactly in the next 
section.  
This entropy is again proportional to area, and correctly gives Bekenstein Hawking entropy.

\subsection{Combinatorial Counting}
\label{count}

In this section, we consider three enumerative problems.  First is the well-understood 
question of 
determining $d$, as defined in (\ref{deg}), the number of 
different ways there are to choose a positive integer $n$ and spins $j_i$ for $1\leq i \leq n$ subject to
the constraint that 

\begin{equation}\sum_i (j_i+1/2) = A.\label{con}\end{equation}
This constraint is due to the bits of semiclassical areas induced by the edges which have to add up to the total area
$A$ of any macroscopic surface measured in units of Planck length squared.
Secondly, we approximate $\frac1d\sum_{j_i}\ln(\prod_i (2j_i+1))$, where the sum is taken over all assignments of 
$n$ and $j_i$ satisfying~(\ref{con}).  We show that this sum is bounded by a constant not depending on $A$.

Finally, we give an exact calculation of $\ln\sum_{j_i}\prod_i (2j_i+1)$ which is required for the solution 
of (\ref{ent2}). 

\medskip
For the duration of this section, we write $a_i$ for $2(j_i+1/2)=2j_i+1$.  The requirement that the $j_i$ be
non-negative half-integers means that the $a_i$ are required to be positive integers.  We also write
$N$ for $2A$.  

The first problem, then, is how many ways to choose $n$ and $a_i$ for $1\leq i\leq n$ 
positive integers, such that $\sum_i a_i=N$.  (These are called {\it compositions} of $N$ in the 
combinatorics literature.)
To solve this, imagine that we have a real line of length $N$,
marked off into $N$ equal pieces by $N-1$ perpendicular lines.  If we choose any subset of the lines as positions to cut the
real line, and then consider (in order) the lengths of the smaller pieces obtained, we will obtain a possible 
assignment of values for $a_1,\dots,a_n$, where $n$ is the number of pieces.  Thus, our problem is equivalent
to asking how many ways we can choose a subset of the $N-1$ lines to cut.  Each line is either chosen or not
chosen, so the total number of compositions of $N$ is $d=2^{N-1}$.

Now we consider the second problem.  
\begin{equation}\sum_{a_i}\ln(\prod_i (a_i))=\sum_{k=1}^N \mu(k)\ln (k),\end{equation}
where $\mu(k)$ is the total number of times $k$ appears as a part, among all the $d$ compositions of $N$. 

Clearly, $\mu(N)=1$, since there is only one composition in which $N$ appears as a part, namely $N=N$.  
To understand $\mu(k)$ for $k<N$, 
consider $\mu_r(k)$, where $\mu_r(k)$ is the number of times $k$ appears as a summand
in one of the $d$ compositions, such that the total of the summands appearing before $k$ is exactly $r$.  

Thus, $\mu_0(k)$ counts the number of compositions of $N$ where the first part is $k$.  
Clearly, by our solution to the first problem, $\mu_0(k)=2^{N-k-1}$.  
Also, $\mu_{N-k}(k)$ counts the number of compositions of $N$ whose final part is $k$, and thus, by symmetry,
$\mu_{N-k}(k)=2^{N-k-1}$.  

Now consider $\mu_r(k)$ for $0<r<N$.  We see that it counts compositions of $N$ which consist of a composition of 
$r$, followed by a part of size $k$, followed by a composition of $N-k-r$.  Thus, 
\begin{equation}\mu_r(k)=2^{r-1}2^{N-k-r-1}=2^{N-k-2}.\end{equation}
Summing, we find that for any $1\leq k <N$, 
\begin{equation} \mu(k)=\sum_{r=0}^{N-k}\mu_r(k)= (N-k+3)2^{N-k-2} \end{equation}

Thus, the quantity we are interested in is given by 
\begin{equation}
\frac{\left(\sum_{k=0}^{N-1} (N-k+3)2^{N-k-2} \ln(k)\right) + \ln (N)}{2^{N-1}}=
\left(\sum_k 2^{-k-1}(N-k+3)\ln(k)\right)+2^{-N+1}\ln(N).
\label{sup}
\end{equation}

The negative 
exponential predominates here, with the result that as $N$ goes to infinity, the sum stays bounded.  

Now we consider the third problem.  We want to understand

$$P_N = \sum_n \sum_{a_1+\dots+a_n=N} \prod_i a_i.$$

The second sum is over all compositions of $N$.  
To help us understand $P_N$, let us fix $n$ and write
\be 
P_{N,n}=\sum_{a_1+\dots+a_n=N} \prod_i a_i.
\label{def}
\ee
We encode the values $P_{N,n}$ in a {\it generating function}:
\be f_n(x)=\sum_{N=0}^\infty P_{N,n}x^N.\ee
Now we claim that $f_n(x)=(x+2x^2+3x^3+4x^4+\dots)^n$.  Compare $x^N$ coefficients on both sides.  The $x^N$
coefficient on the lefthand side is $P_{N,n}$ by assumption.  On the righthand side, contributions to 
$x^N$ will be obtained by choosing a term from each of the $n$ factors, such that the exponents of the chosen
terms add to $N$.  Thus, this choice amounts to a choice of a composition of $N$.  Now we see that the 
coefficients were chosen precisely so that the weight corresponding to a given composition is the product of the 
parts, and thus the coefficient of $x^N$ on the righthand side is also $P_{N,n}$, as desired.   

So 
$f_n(x)=(x+2x^2+3x^3+\dots)^n=\left(\frac x{(1-x)^2}\right)^n$.
Summing, we find that \be f(x)=\frac{\frac x{(1-x)^2}}{1-\frac x{(1-x)^2}}=\frac x{1-3x+x^2}.\ee

Thus, we have obtained a generating function for $P_N$.  Note that this is essentially the same calculation
as was carried out in \cite{alek}; their partition function $Z$ is our generating function 
$f(x)$ once one sets $e^{-\beta}=x$.  

Write $1-3x+x^2=(1-\alpha x)(1 - \beta x)$, where 
$$\alpha= \frac{3+\sqrt{5}}2,\ \beta =\frac{3-\sqrt{5}}2.$$

Expanding in partial fractions,
we get
\begin{eqnarray}
f(x)&= & \frac{1/\sqrt{5}}{1-\alpha x} -\frac{1/\sqrt{5}}
{1-\beta x}\\
&= & \frac{1}{\sqrt{5}} \sum_N \left(\left(\frac{3+\sqrt{5}}{2}\right)^N
-\left(\frac{3-\sqrt{5}}{2}\right)^N\right)x^N\end{eqnarray}

Therefore, $$P_N=\frac{1}{\sqrt 5}\left(\left(\frac{3+\sqrt{5}}{2}\right)^N
-\left(\frac{3-\sqrt{5}}{2}\right)^N\right).$$

Thus, the quantity we are interested in, $\ln(P_{2A})$, is given by
\be
\ln(P_{2A})=\ln\left(\frac{1}{\sqrt 5}\left(\left(\frac{3+\sqrt{5}}{2}\right)^{2A}
-\left(\frac{3-\sqrt{5}}{2}\right)^{2A}\right)\right).
\label{are}
\ee

Thus the above would the the degeneracy of the area of any macroscopic surface with total
area $A$.

\subsection{Entropy}
Now, after obtaining exact expressions for the various ways of counting the horizon, we now
specialise to the determination of entropy.
The basic formalism is the same: we are counting the number of ways to induce the horizon area
by arbitrary graphs, in the semi-classical limit. 
The horizon area is given by a sum of bits of areas which equal some integer given by $(j_{\rm cl} + \frac12)t$
where $j_{\rm cl}$ is the spin of the edge inducing the area.

 In most of the previous countings of entropy, the spins were all set equal to $1/2$, 
and a justification for that appears in \cite{spin}. However, here, they would correspond to
the choice of a very symmetric graph, where all the spins induce the horizon with precisely
an area $t$. The entropy counting would thus be trivial and equal 
\be
S^{1}_{BH} = \frac{A_H}{l_p^2} \ln 2
\label{equal}
\ee
Now, this is not exactly the same as the Bekenstein-Hawking entropy as the constant is not
$1/4$. This problem is solved as in \cite{abck}, by realising that the area apectrum is modified
by the Immirzi parameter in different quantisation sectors of the theory. Thus the area
bit is $(j+1/2)t \beta$, and hence the entropy term gets modified now by the requirement that the
integers have to add up to area divided by the Immirzi parameter. Thus the constant $N= 2A/\beta$ 
(in units of Planck length squared).  
The Immirzi parameter would be $4 \ln 2$, though not exactly equal to the
value obtained from previous calculations of entropy \cite{abck}. In fact, what is being
identified here is a problem: the lack of universality of the `Immirzi parameter'. 
Every time  one gives a counting of black hole entropy within the formalism of
loop quantum gravity, one discovers a new value of the same parameter.
Though by definition, the parameter is an ambiguity in quantisation, these
`theoretical' experiments do not actually fix the ambiguity in any way. Firstly, given the
level of theoretical knowledge, it is difficult to determine the correct entropy
calculation. However, the universality of Bekenstein-Hawking entropy as being
proportional to the area of the horizon is true for all countings. This law
will hopefully be verifiable experimentally, by actual observations.
Note also in the above case (\ref{equal}), the counting is exact, and there are no corrections
semi-classically to the area law. There will be quantum fluctuations which will correct the
classical space-time, however that need not be proportional to area and would be one order
higher in the semi-classical parameter $t$.

Now, suppose the number of spins were $i=1..n$. Then the density matrices would be
 $\prod^{n}_{i=1} (2j_i+1) \times \prod^{n}_{i=1} (2j_i+1)$ matrices and hence the 
entropy would be of the form:
\begin{eqnarray}
S_{BH}&=& \sum_i\ln(2j_i+1)\\
&=& \ln(\prod_i (2j_i+1)).
\end{eqnarray}
In addition there will be the constraint that the spins sum up to the total area
$\sum_i(j_i+1/2)t= A_H/l_p^2$.
For a fixed graph, where the number of edges is fixed, the classical area $P_H$ would thus be fixed
and hence the spin counting $i=1..n$ also fixed to be a certain number.  However, what is now different
from the all spins equal case, is that area is not equally distributed, and hence in some sense
one is summing over inequivalent graphs but with the same number of edges crossing the horizon. This
would correspond to the diagram of the previous section. This entropy would correspond
to the log of $P_{N,n}$ of equation (\ref{def}). To extract the coefficient of $P_{N,n}$ from
$f_n$, 
\be
f_n(x)= x^n(1-x)^{-2n}= x^n \sum_{i=0}^{\infty} \left(\begin{array}{c}-2n\\i\end{array}\right)(-1)^i x^i.
\ee
Hence, the coefficient of $x^N$ will be
\be
P_{N,n}=\left(\begin{array}{c}-2n\\N-n\end{array}\right)(-1)^{N-n}=\frac{2n(2n+1)......(n+N-1)}{(N-n)!}
\ee
Now, the n can be absolutely arbitrary, but that will create a dependence of the entropy on this
unknown parameter. However, to gauge some understanding of the entropy we take the number of spins to be
$N/2$, so as to compare with the case where all the spins were set to $1/2$.

Clearly,
\be
P_{N,N/2} = \frac{N(N+1)...(\frac{3N}{2} -1)}{\frac N2!}= \frac{2}{3N}\frac{\frac{3N}{2}!}{(N-1)!\frac N2 !}
\ee
Using the Stirling's formula, $r!= r^r e^{-r} \sqrt{2 \pi r}$, for large $r$, and one obtains 
the following:
\bea
P_{N,N/2} & = & \frac{2}{3N}\left(\frac{3N}{2}\right)^{3N/2}e^{-3N/2+ N/2 +N} \frac{N}{2}^{-N/2} (N-1)^{-N+1} \frac{\sqrt{ \pi (3N)}}{\sqrt{2\pi N}\sqrt{\pi N}} \nn \\
&=& \frac{2\sqrt{3}}{3\sqrt{2 \pi} N^{1/2}}  3^{3 N/2} 2^{-N} + O(\frac1 N) 
\eea
The entropy or log of this quantity is then
\be
S_{BH} = N(\frac32\ln3 - \ln 2) - \frac12\ln N  +\frac12\ln(\frac{2\pi}{3})  + {\rm further  \ \ corrections}
\ee
Now, $ N = 2 \frac{A_H}{ \beta l_p^2}$, and thus we are observing a particular value of the Immirzi parameter, and a
correction to Bekenstein-Hawking \cite{bek,hawk} entropy which is proportional to log of area, already observed.
Of course, here, the area spectrum is different, and the counting is exact. Note that this entropy is different
from the entropy when all the spins were symmetrically distributed (\ref{equal}).

\subsection{Sum over all spins}
While trying to understand the sum over all possible graphs, and the corresponding entropy, one takes,
as a first case, the tensor sum of density matrices for individual graphs.
The entropy is then given by (\ref{deg}). The combinatorial counting for $d$ has been done, and the
correction term is exponentially supressed as given in (\ref{sup}). Thus the entropy is
\be
S_{BH} \approx (N-1)\ln 2 + \exp(-(N-1)\ln 2) \ln N
\ee
Note that $N= 2A_H/l_p^2 \beta$ and there is indeed a $\ln A$ term, however, that is multiplied by a term
which decreases exponentially with area. If one expanded the exponential, then the first term 
would be a log correction term, however, this would be true only for small enough areas.

Now, we take the physically interesting case where density matrix is ab-initio derived from the generalised
coherent state. This
is given by the expression for $P_{2A}$ in equation (\ref{are}). This is surprisingly similar to the
one obtained in \cite{rov}, where the counting was done for the number of integer partitions of a given area.
The entropy is thus
\bea
\ln P_{2A}& =& 2 \frac{A_H}{\beta l_{pl}^2} \ln (\frac{3 +\sqrt{5}}{2}) + \ln(1- (6.854)^{-2 A_H/\beta l_{pl}^2}) - \frac12\ln 5\\
& = & 2 \frac{A_H}{\beta l_{pl}^2}\ln(\frac{3 + \sqrt{5}}{2}) + (6.854)^{-2 A_H/\beta l_{pl}^2} + ...
\eea
Clearly, the first term is the Bekenstein-Hawking entropy, but the corrections now are exponentially
decreasing in area. As area increases, the corrections go to zero. The Immirzi parameter obtained here
is precisely proportional to the one obtained in \cite{alek}, which is not surprising as the
counting here is for the same object. This new observation regarding correction to entropy might have
a origin from quantum information theory \cite{info}.
The correction to the entropy is a completely new result here, though the implications of this are 
not very clear. 
Classical laws of black hole mechanics are usually supplemented with Hawking temperature to 
get the exact 1/4 factor. But this can have semi-classical corrections and considerably
modify our interpretations.

\section{Conclusions}
We defined generalised coherent states over a tensor sum of Hilbert spaces labelled by minimal
graphs and their corresponding classical labels. We derived the density matrix from the
corresponding generalised coherent state and showed that the entropy is proportional to area.
The corrections to entropy are exponentially decreasing with area. 
These coherent states are still defined in the kinematic Hilbert space. With the new work
on the Hamiltonian constraint, we hope to be able to contribute to the derivation of the 
coherent state in the constraint physical Hilbert space. However, since in the limit $t\rightarrow 0$,
the results predicted by those physical coherent states have to agree with the classical results,
we do not expect much deviation from the conclusions of our paper.

{\bf Acknowledgements:} We would like to thank S. Das, B. Dittrich, J. Gegenberg and S. Vardarajan for useful discussions. This research is funded
by NSERC and research funds of University of New Brunswick.

\end{document}